\documentclass[final,5p,12pt,twocolumn,number,sort&compress]{elsarticle}

\usepackage{amsmath}
\usepackage[utf8]{inputenc}
\usepackage{graphicx}
\usepackage{dcolumn}
\usepackage{bm}
\usepackage{hyperref}
\hypersetup{
    colorlinks=true,
    allcolors=blue,
}
\usepackage{xcolor}
\usepackage[normalem]{ulem}
\usepackage{cancel}

\newif\ifrevision
\revisionfalse

\ifrevision
  \newcommand{\add}[1]{\textcolor{blue}{#1}}
  \newcommand{\del}[1]{\textcolor{red}{\sout{#1}}}
  \newcommand{\replace}[2]{\del{#1}\add{#2}}
  \newcommand{\note}[1]{\textcolor{teal}{[Note: #1]}}
\else
  \newcommand{\add}[1]{#1}
  \newcommand{\del}[1]{}
  \newcommand{\replace}[2]{#2}
  \newcommand{\note}[1]{}
\fi

\newcommand{\eq}[1]{Eq.~\eqref{#1}}

\begin{document}
\begin{frontmatter}
% \title{A Novel Solution to Bumblebee Gravity}
\title{ Static Spherical Vacuum Solution to Bumblebee Gravity with Time-like VEVs}
\author[inst1]{Hao Li}
 \ead{haolee@cqu.edu.cn}
 %\affiliation{Department of Physics, Chongqing University, Chongqing 401331, P.R. China}

\author[inst1]{Jie Zhu\texorpdfstring{\corref{cor1}}{}}
  \ead{jiezhu@cqu.edu.cn}
  \cortext[cor1]{Corresponding author}
   \affiliation[inst1]{
   organization={Department of Physics and Chongqing Key Laboratory for Strongly Coupled Physics, Chongqing University},
   city={Chongqing 401331},
   country={P. R. China}
   }
   % \affiliation[inst2]{
   % organization={Department of Physics and Chongqing Key Laboratory for Strongly Coupled Physics, Chongqing University,},
   % city={Chongqing 401331},
   % country={P. R. China}
   % }
   %{Department of Physics, Chongqing University, Chongqing 401331, P.R. China}

\date{\today}

\begin{abstract}
    The static spherical vacuum solution in a bumblebee gravity model where the bumblebee field \(B_\mu\) has a one-component time-like vacuum expectation value \(b_\mu\) is studied.
    We show that in general curved space-time solutions are not allowed and only the Minkowski space-time exists.
    However, it is surprising that \add{two} non-trivial solutions can be obtained so long as a unique condition for the vacuum expectation \(b^2\equiv -b^\mu b_\mu=2/\kappa\), where \(\kappa=8\pi G\), is satisfied.
    \add{One of the solutions contains a naked singularity, while the other exhibits features analogous to a confining potential.}
    We argue that naturally these solutions are not stable since quantum corrections would invalidate the likely numerical coincidence, unless there are some unknown \emph{fine-tuning} mechanisms preventing any deviation from this condition.
    Nevertheless, 
    the naked singularities and the photon sphere of these novel but peculiar solutions are discussed,
    and we show that the extremal Reissner-Nordstr{\"o}m solution is a limit of one of our solutions.
\end{abstract}


\end{frontmatter}
%\maketitle

\section{Introduction}

General Relativity~(GR) and the Standard Model~(SM) of particle physics are the most successful theories describing all four fundamental forces of nature.
However, there are theoretical tensions between GR and SM, and to reconcile them, several candidates of quantum gravity~(QG) theories have already been proposed.
Generally, the onset of significant effects of QG is expected to happen at the Planck scale~(\(E_{Pl}\sim 10^{19}~\text{GeV}\)), which is far beyond our reach for current experiments.
Although direct detection of QG effects seems to be unlikely at present, it is suggested that there exists the possibility that certain kinds of remnant signals of QG could be observed at energy scales much lower than the Planck scale.
One of such signals is the violation of Lorentz invariance.

In recent decades, increasing interest has been directed toward possible violations of Lorentz symmetry, driven by attempts to formulate a consistent theory of quantum gravity and to understand potential deviations from GR at high energies. Among various approaches, effective field theories incorporating spontaneous Lorentz violation have proven especially fruitful. In this context, Bumblebee gravity has emerged as a minimal yet nontrivial extension of GR, wherein a vector field acquires a nonzero vacuum expectation value (VEV), leading to spontaneous breaking of local Lorentz invariance and diffeomorphism invariance in a controlled manner~\cite{Kostelecky:1988zi,Kostelecky:2003fs,Bluhm:2004ep,Bailey:2006fd,Maluf:2013nva}~\footnote{\add{An alternative framework to incorporate Lorentz invariance violation in gravity is introduce a background Kalb-Ramond field~\cite{al-Badawi:2024pdx,Mangut:2025gie,Al-Badawi:2025ejf}.}}.

The Bumblebee model typically introduces a vector field $B_{\mu,}$ coupled non-minimally to gravity and
governed by a potential $V(B_\mu B^\mu\pm b^2)$, which determines the vacua of the theory at the classical level.
The vacua, determined by the vacuum expectation \(\langle B^\mu\rangle\) such that \(V|_{B_\mu=\langle B_\mu\rangle}\), do not transform as scalars, thus signaling the spontaneously breaking of Lorentz symmetry.
% enforces the constraint $B_\mu B^\mu=\pm b^2$ at the level of field equations.
This spontaneous Lorentz symmetry breaking results in modifications to the Einstein field
equations, leading to potentially observable signatures in gravitational phenomena. 
In previous studies, an exact Schwarzschild-like solution in bumblebee gravity is proposed~\cite{Casana:2017jkc}.
Also, an exact Kerr-like solution is found~\cite{Ding:2019mal}.
But for both of the solutions, the configurations of the bumblebee field only admit space-like VEVs, i.e., $\langle B^\mu B_\mu\rangle > 0$.
Similar results are also obtained and analyzed in detail within metric-affine bumblebee models~\cite{Filho:2022yrk,AraujoFilho:2024ykw,AraujoFilho:2025hkm}, \add{models with cosmological constant~\cite{Valtancoli:2023kdy}, higher-dimensional dS/AdS spacetimes~\cite{Uniyal:2022xnq}, models with topological defects~\cite{Gullu:2020qzu}, and models with generalized uncertainty principle modifications~\cite{Kanzi:2019gtu,Karmakar:2023mhs}.
Meanwhile, phenomenology of bumblebee gravity is also investigated~\cite{Kanzi:2021cbg,Karmakar:2023mhs,Mangut:2023oxa,Neves:2024ggn,Uniyal:2022xnq}.
}
The purpose of this work is to investigate the solution to the Bumblebee gravity model with a time-like vacuum expectation value of the bumblebee field.

This literature is organized as follows.
In Sec.~\ref {sec:2}, we briefly introduce the action and the equation of motion for bumblebee gravity.
In Sec.~\ref {sec:3}, we try to solve the equations of motion in a static spherical field configuration with the bumblebee field obtaining one-component time-like VEVs, and we find that for general VEVs of this type there is no non-trivial solution for the equations, unless $b=\sqrt{2/\kappa}$.
In Sec.~\ref{sec:4}, we solve the equations when $b=\sqrt{2/\kappa}$, and find two kinds of non-trivial solutions.
Sec.~\ref{sec:5} is the discussion and the summary.
In this literature, we will adopt the metric signature $(- + + +)$ and also all quantities involved are expressed in natural units $(\hbar=c=1)$.

\section{Brief Introduction of Bumblebee gravity}\label{sec:2}
The action of Bumblebee gravity can be expressed as~\cite{Casana:2017jkc}~\footnote{We have omitted additional terms proportional to the cosmological constant \(\Lambda\), \(B^\mu B_\mu R\), \(\nabla_\mu B_\nu\nabla^\mu B^\nu\) and \((\nabla_\mu B^\mu)^2\)}
\begin{align}
    S=&\int d^4x\sqrt{-g}\nonumber \\ \times&\left(\frac{1}{2\kappa}R+\frac{\xi}{2\kappa}B^\mu B^\nu R_{\mu\nu}-\frac{1}{4}B^{\mu\nu}B_{\mu\nu}-V\right)\nonumber\\ +&S_\mathrm{m},\label{action}
\end{align}
where $g$ is the determinant of the metric $g_{\mu\nu}$, the constant $\kappa\equiv8\pi G$ with $G$ being the gravitational constant, $S_{\mathrm{m}}$ represents the action for matter fields of no interest in this work, $B_\mu$ is the bumblebee field, and the field strength tensor is $B_{\mu\nu}=\partial_{\mu}B_{\nu}-\partial_{\nu}B_{\mu}$.
In bumblebee theories, the potential $V$ is selected to provide a non-vanishing VEV for $B_\mu$, and could have the following general functional form
\begin{equation}
    V\equiv V(B^\mu B_\mu + s b^2),\label{potential}
\end{equation}
where $b$ is a positive real constant, and $s=\pm 1$ to determine whether the expection value of $B_\mu$ is time-like or space-like. 
In the literature, it is usually assumed that \(V\) has (at least one of) its minimum/maximum at \(0\), thus~\footnote{The condition for \(V''(0)\) plays no role in this work.}
\begin{equation}
    V(0)=0,\ \text{and}\ V'(0)=0.\label{vacuumcondition}
\end{equation}
The VEV of the bumblebee field is determined when $V(B^\mu B_\mu + s b^2)=0$, implying that
\begin{equation}
    B^\mu B_\mu+s b^2=0, 
\end{equation}
The above equation provides a non-null vacuum expectation value
\begin{equation}
    \langle B^\mu\rangle=b^\mu,
\end{equation}
where $b_\mu b^\mu + s b^2=0$.

We are interested in the vacuum solution, namely $(T_m)_{\mu\nu}=0$. 
In this work, we consider the case that $B_\mu$ admits VEV as $b_\mu$ and there is no cosmological constant, so for $V$ we have
\begin{equation}
    \begin{aligned}
     \left.V\right|_{B_\mu=b_\mu}&=0,\\
    \left.\frac{dV}{d(B^\lambda B_\lambda)}\right|_{B_\mu=b_\mu}&=0.
    \end{aligned}
\end{equation}
Provided with the above condition for $V$ and $B_\mu$ replaced exactly by its VEV $b_\mu$, the equation of motions for $g_{\mu\nu}$ is
\begin{equation}
    R_{\mu\nu}-\frac{1}{2}R\,g_{\mu\nu}-\kappa T^B_{\mu\nu}=0,
\end{equation}
where 
\begin{equation}
    \begin{aligned}
    &T_{\mu\nu}^{B}=-B_{\mu\alpha}B_{\nu}^{\alpha}-\frac{1}{4}B_{\alpha\beta}B^{\alpha\beta}g_{\mu\nu}
    % -Vg_{\mu\nu}+2V^{\prime}B_{\mu}B_{\nu}
    \\&+\frac{\xi}{\kappa}\left[\frac{1}{2}B^{\alpha}B^{\beta}R_{\alpha\beta}g_{\mu\nu}-B_{\mu}B^{\alpha}R_{\alpha\nu}-B_{\nu}B^{\alpha}R_{\alpha\mu}\right]\\&+\frac{1}{2}\nabla_{\alpha}\nabla_{\mu}\left(B^{\alpha}B_{\nu}\right)+\frac{1}{2}\nabla_{\alpha}\nabla_{\nu}\left(B^{\alpha}B_{\mu}\right)\\&-\frac{1}{2}\nabla^2\left(B_\mu B_\nu\right)-\frac{1}{2}g_{\mu\nu}\nabla_\alpha\nabla_\beta\left(B^\alpha B^\beta\right).
    \end{aligned}
\end{equation}
Take the trace of the equation for $g_{\mu\nu}$, we get $R=-\kappa T^B$, and thus the above equation can be expressed as
\begin{equation}
    R_{\mu\nu}-\kappa(T^B_{\mu\nu}-\frac{1}{2}g_{\mu\nu}T^B)=0. \label{eq:gt}
\end{equation}
For $B_\mu$, the equation of motion is
\begin{equation}
    % \frac{\xi}{\kappa}B^\mu R_{\mu \nu}+\nabla_\mu\nabla^\mu B_\nu-\nabla_\mu\nabla_\nu B^\mu=0. 
    \frac{\xi}{\kappa}B^\mu R_{\mu \nu}+\nabla^{\mu}B_{\mu\nu}=0.\label{eq:b}
\end{equation}

\add{
As discussed in Ref~\cite{Ji:2024aeg}, the static spherical vacuum solution to Bumblebee Gravity can be classified into two classes. For the static spherical field configuration \(B_\mu=b_\mu=(b_t(r), b_r(r), 0, 0)\), the only non-vanishing components of field strength tensor $b_{\mu\nu}=\partial_\mu b_\nu-\partial_\nu b_\mu$ are \(b_{rt}=-b_{tr}=b_t'(r).\)
Rewriting \eq{eq:b} as
\begin{equation}
    \frac{1}{\sqrt{-g}}\partial_{r}(\sqrt{-g}b^{\bar{\mu}r})-\frac{\xi}{\kappa}R^{\bar{\mu}\bar{\mu}}b_{\bar{\mu}}=0,
\end{equation}
where there is no summation for the index \(\bar{\mu}=t,r,\theta,\phi\),
we can easily see that the $r-$~component of \eq{eq:b} is
\begin{equation}
    \frac{\xi}{\kappa} R^{rr}b_r=0.\label{eq:Rrrbr}
\end{equation}
\eq{eq:Rrrbr} shows that the spherically symmetric vacuum solutions in the bumblebee model can be classified into two \emph{disjoint} classes:
\begin{itemize}
    \item Class I: $b_r \equiv 0$,
    \item Class II: $R_{rr} \equiv 0$.
\end{itemize}
For the space-like VEVs, $b_r$ cannot be identically zero, so the solutions are classified as belonging to Class II,
whereas in the time-like cases, $b_r\equiv 0$ can be achieved.
In this work, we consider only the Class I solutions.
Regarding the Class II solutions in the time-like configuration, one can consult Ref.~\cite{Zhu:2025fiy}.
}

In fact, \eq{eq:gt} and \eq{eq:b} are the same equations of motion for a massless vector field non-minimally coupled to gravity, i.e., action \eq{action} with $V\equiv0$.
But in the bumblebee model, one more constraint $B_\mu B^\mu+sb^2=0$ is added.
Luckily, the exact Schwarzschild-like solution with space-like VEV for $B_\mu$, explored in Ref.~\cite{Casana:2017jkc}, satisfies all of the equations of motion and the constraint.
In general, the existence of a solution that satisfies all of the equations and the constraint simultaneously is unlikely.
And this is what we will discuss in the next section.

\section{Static spherical Vacuum solution for bumblebee gravity with general time-like VEVs}\label{sec:3}

% \subsection{General cases}

% The static spherical solution for bumblebee gravity with space-like VEV is explored in Ref.~\cite{Casana:2017jkc}, and here
Here we consider the time-like case and fix \(s=1\) in Eq.~\eqref{potential} in the following. 
According to the Birkhoff theorem, we adopt the following metric
\begin{equation}
    g_{\mu\nu}=\mathrm{diag}\left(-e^{2\alpha},e^{2\beta},r^2,r^2 \sin^2\theta\right),
\end{equation}
where $\alpha$ and $\beta$ are functions of $r$.
\replace{We consider a one-component time-like ground $b_\mu$ as}{For the Class I solutions, the $b_\mu$ is}
\begin{equation}
    b_\mu=(b_t(r),0,0,0).
\end{equation}
By $b_\mu b^\mu+b^2=0$, we have
\begin{equation}
    b_t(r)=b e^{\alpha(r)}.
\end{equation}

Substitute the above equations into \eq{eq:gt} and \eq{eq:b} and define $\ell=\xi b^2$, we have the following expressions to be zero for the gravity sector
\begin{subequations}
\begin{equation}
\begin{aligned}
    \rm{EQ}_{tt}=&+(\ell-2)r \alpha''(r)+(\kappa b^2-2+\ell)r \alpha'(r)^2\\
             &-(\ell-2)\alpha'(r)\left(r\beta'(r)-2 \right),\\
\end{aligned}
\label{eqg:a}
\end{equation}
\begin{equation}
\begin{aligned}
    \rm{EQ}_{rr}=&-(\ell+2)r \alpha''(r)+(\kappa b^2-2-\ell)r \alpha'(r)^2\\
            &+4\beta'(r)+\alpha'(r)\left((\ell+2)r\beta'(r)-2\ell\right),\\
\end{aligned}
\label{eqg:b}
\end{equation}
\begin{equation}
\begin{aligned}
\rm{EQ}_{\theta\theta}=&-\ell r^2\alpha''(r)-(\kappa b^2+\ell)r^2\alpha'(r)^2\\
    &+r\alpha'(r)(\ell r \beta'(r)-2(1+\ell))\\
    &+2(r\beta'(r)+e^{2\beta(r)}-1),\\
\end{aligned}
\label{eqg:c}
\end{equation}
\begin{equation}
\begin{aligned}
\rm{EQ}_{\phi\phi}=& \sin^2\theta\cdot \rm{EQ}_{\theta\theta},
\end{aligned}
\label{eqg:d}
\end{equation}
\label{eqg}
\end{subequations}
and
\begin{equation}
\begin{aligned}
\rm{EQ}^B_{t}=&-(\kappa b^2-\ell)r\alpha''(r)+\ell r \alpha'(r)^2\\
        &+(\kappa b^2-\ell)\alpha'(r)(r\beta'(r)-2)
\end{aligned}
\label{eqB}
\end{equation}
for $B_\mu$ sector.
Notice that by applying 
\begin{equation}
    (\kappa b^2-\ell)\rm{EQ}_{tt}+(\ell-2)\rm{EQ}^B_t=0,
\end{equation}
we have the following relation
\begin{equation}
    \kappa b^2 (\kappa b^2-2)r \alpha'(r)^2=0.\label{maineq}
\end{equation}
So for general cases with $b \neq 0$ and $b\neq\sqrt{2/\kappa}$, we have $\alpha'(r)=0$. Substitute it into \eq{eqg:b}, we have $\beta'(r)=0$.
Thus, there is no non-trivial static spherical \add{Class I} solution for bumblebee gravity with general time-like VEVs. This is one of the main results of this work.
Next we focus on solutions with the condition \(b=\sqrt{2/\kappa}\) satisfied.

\section{Solution for bumblebee gravity with VEV \texorpdfstring{$b=\sqrt{2/\kappa}$}{}}\label{sec:4}

\add{In the case of $b=\sqrt{2/\kappa}$, $\ell=\xi b^2=\frac{2\xi}{\kappa}$, and \eq{eqg} and \eq{eqB} reduce to 
\begin{subequations}
\begin{equation}
\begin{aligned}
    \rm{EQ}_{tt}=&(\ell-2)r \alpha''(r)+\ell r \alpha'(r)^2\\
             &-(\ell-2)\alpha'(r)\left(r\beta'(r)-2 \right),\\
\end{aligned}
\label{eqg:a1}
\end{equation}
\begin{equation}
\begin{aligned}
    \rm{EQ}_{rr}=&-(\ell+2)r \alpha''(r)-\ell r \alpha'(r)^2\\
            &+4\beta'(r)+\alpha'(r)\left((\ell+2)r\beta'(r)-2\ell\right),\\
\end{aligned}
\label{eqg:b1}
\end{equation}
\begin{equation}
\begin{aligned}
\rm{EQ}_{\theta\theta}=&-\ell r^2\alpha''(r)-(2+\ell)r^2\alpha'(r)^2\\
    &+r\alpha'(r)(\ell r \beta'(r)-2(1+\ell))\\
    &+2(r\beta'(r)+e^{2\beta(r)}-1),\\
\end{aligned}
\label{eqg:c1}
\end{equation}
\label{eqg1}
\end{subequations}
and
\begin{equation}
\begin{aligned}
\rm{EQ}^B_{t}=&(\ell-2)r\alpha''(r)+\ell r \alpha'(r)^2\\
        &-(\ell-2)\alpha'(r)(r\beta'(r)-2).
\end{aligned}
\label{eqB1}
\end{equation}
}
\add{
Now we have 
\begin{equation}
\begin{aligned}
0=&\frac{r}{4}\left(\rm{{EQ}_{tt}-\rm{EQ}_{rr}} \right)+\frac{1}{2}\rm{EQ}_{\theta\theta}\\
=&e^{2\beta(r)}-(1+r \alpha'(r))^2,\label{eq:rba}
\end{aligned}
\end{equation}
which gives 
\begin{equation}
    \beta(r)=\ln\left(1+r \alpha'(r)\right)\label{eq:rba1},
\end{equation}
and 
\begin{equation}
    \beta'(r)=\frac{\alpha'(r)+r \alpha''(r)}{1+r \alpha'(r)}.\label{eq:rba2}
\end{equation}
}
% \del{
% Notice that by applying 
% \begin{equation}
%     \frac{r}{4}\left(\rm{{EQ}_{tt}-\rm{EQ}_{rr}} \right)+\frac{1}{2}\rm{EQ}_{\theta\theta}=0,
% \end{equation}
% and replacing $b$ with $\sqrt{2/\kappa}$, we have
% \begin{equation}
%     e^{2\beta(r)}-(1+r \alpha'(r))^2=0.\label{eq:rba}
% \end{equation}
% }
\replace{Substitute \eq{eq:rba} into \eq{eqg} and \eq{eqB}, }{Substitute \eq{eq:rba1} and \eq{eq:rba2} into \eq{eqg1} and \eq{eqB1},}
surprisingly, we can find that all of the equations become the same equation as
\begin{equation}
\begin{aligned}
    (\ell-2)r\alpha''(r)&+\ell r^2 \alpha'(r)^3+2(\ell-1)r\alpha'(r)^2\\
    &+2(\ell-2)\alpha'(r)=0.\label{eq:alpha}
\end{aligned}
\end{equation}
So in the case of $b=\sqrt{2/\kappa}$, static spherical vacuum solutions can exist, if we could find a solution of \eq{eq:alpha}.

\subsection{Solutions with constant \texorpdfstring{$\beta$}{}}\label{sec:Constantb}

In this case, from \eq{eq:rba}, $\alpha$ can be expressed as
\begin{equation}
    \alpha(r)=A \ln(\frac{r}{R_0}),
\end{equation}
where $R_0$ is a constant.
Substituting it into \eq{eq:alpha}, we can obtain three solutions as $A=0$, $A=-1$, and $A=\frac{2}{\ell}-1$.
The case $A=0$ is the flat Minkowski space-time.
The case $A=-1$ is ruled out because $g_{rr}=(1+r \alpha'(r))^2=0$.
When $A=\frac{2}{\ell}-1$, we can get
\begin{equation}
\begin{aligned}
g_{tt}&=-e^{2\alpha}=-\left(\frac{r}{R_0}\right)^{2(2/\ell-1)},\\
g_{rr}&=(1+r \alpha'(r))^2=\frac{4}{\ell^2},\\
b_t&=\sqrt{\frac{2}{\kappa}}e^{\alpha}=\sqrt{\frac{2}{\kappa}}\left(\frac{r}{R_0}\right)^{2/\ell-1}.
\end{aligned}
\end{equation}
Thus the metric of the solution is 
\begin{equation}
    g_{\mu\nu}=\mathrm{diag}\left(-\left(\frac{r}{R_0}\right)^{\frac{2(2-\ell)}{\ell}},\frac{4}{\ell^2},r^2,r^2 \sin^2\theta\right).\label{eq:sol1}
\end{equation}
When $\ell=2$, i.e., $\xi=\kappa$, we can see that the above solution degenerates into the flat Minkowski spacetime.
And of course, this solution does not admit the limit $\ell\rightarrow 0$.
The Kretschmann scalar of this solution is
\begin{equation}
    K=R_{\mu\nu\rho\sigma}R^{\mu\nu\rho\sigma}=\frac{(\ell-2)^2(7\ell^2-4\ell+8)}{4r^4},\label{eq:kretschmann1}
\end{equation}
which means that $r=0$ is a singularity of this solution~\footnote{Indeed the other numerator factor \(7\ell^2-4\ell+8=0\) has no real root, therefore the singularity cannot be cured by carefully choosing the value of \(\ell\) besides \(\ell=2\).}.
So this solution admits a naked singularity located at $r=0$.
The Einstein tensor for the solution \eq{eq:sol1} is
\begin{align}
    G^\mu_\nu=\mathrm{diag}\big(&(\ell^2-4)A,-(\ell-2)^2A,\nonumber \\ &(\ell-2)^2A,(\ell-2)^2A\big),
\end{align}
where $A=1/4r^2$.
So we can easily see that when $0\leq \ell\leq2$, the metric satisfies the dominant energy condition~(DEC), and consequently the weak, null, and null dominant energy condition~(WEC, NEC, NDEC).
Furthermore, the strong energy condition~(SEC) is also satisfied in this case.

\subsection{Solutions with non-constant \texorpdfstring{$\beta$}{}}

It seems that from \eq{eq:rba} we have two solutions as $1+r \alpha'(r)=\pm e^{\beta(r)}$.
In fact, this appears because of our choice of coordinates,  and both of them provide the same solution.
For simplicity, we only consider the case $1+r \alpha'(r)=- e^{\beta(r)}$.
It is hard to solve \eq{eq:alpha} directly.
Using 
\begin{equation}
    \alpha'(r)=-\frac{1+e^{\beta(r)}}{r}\label{eq:rab}
\end{equation}
and substituting it into \eq{eq:alpha},
we get the equation for $\beta(r)$ as
\begin{equation}
    (\ell-2)r\beta'(r)+(e^\beta+1)(\ell e^\beta+2)=0.\label{eq:beta}
\end{equation}
Let $u(r)=e^{-\beta(r)}$, we have
\begin{equation}
    \frac{1}{r}\frac{dr}{du}=\frac{(\ell-2)u}{(u+1)(2u+\ell)},
\end{equation}
and the solution is
\begin{equation}
    \frac{r}{R_0}=\frac{(2u+\ell)^{\ell/2}}{u+1},\label{eq:solu}
\end{equation}
where $R_0$ is a constant.

Firstly, we consider the case of $\ell\rightarrow 0$ (i.e., $\xi\rightarrow 0$).
Then \eq{eq:solu} gives $u=R_0/r-1$, leading to
\begin{equation}
    g_{rr}=e^{2\beta}=\frac{1}{u^2}=\left(1-\frac{R_0}{r}\right)^{-2}.
\end{equation}
The solution for $\alpha$ is 
\begin{equation}
    \alpha=\ln\left(1-\frac{R_0}{r}\right)+C,
\end{equation}
thus
\begin{equation}
    g_{tt}=-e^{2\alpha}=-\left(1-\frac{R_0}{r}\right)^{2}
\end{equation}
up to an overall constant that can be absorbed into a redefinition of $t$.
The solution for $B_\mu$ is
\begin{equation}
    b_t=\sqrt{\frac{2}{\kappa}}(1-\frac{R_0}{r}),
\end{equation}
and the metric is
\begin{align}
g_{\mu\nu}=\mathrm{diag}\Big(-&\big(1-\frac{R_0}{r}\big)^{2},\big(1-\frac{R_0}{r}\big)^{-2},\nonumber \\
 & r^2,r^2 \sin^2\theta\Big).    
\end{align}
Actually, this solution coincides with the extremal Reissner–Nordstr{\"o}m black hole solution, while the solution for \(B_\mu\) is not the same: in the Reissner-Nordstr{\"o}m solution, \(B_{\mu}=(Q/r,0,0,0)\) which obeys \(B_{\mu}(r)=0\ \text{for }r\to\infty\); in our case, however, the temporal component of \(B_{\mu}\) at the infinity is a non-zero constant \(\sqrt{2/\kappa}\).
\replace{Nevertheless, the difference is merely a consequence of choosing different boundary conditions for \(B_{\mu}\).}{The asymptotic behavior $b_t(r)\to\sqrt{2/\kappa}$ follows from the constraint $b_\mu b^\mu=-b^2;$ we note that an
analogous constant asymptotic value may also appear in the electromagnetic case, as it can be generated
by a gauge transformation of $A_\mu.$}

%% extremal RN: |Q|=\sqrt{G} M, Q^2=G R0^2, M=G R0

For the general cases, from \eq{eq:solu} and \eq{eq:rab}, we have
\begin{equation}
\frac{d \alpha}{d u}=-\frac{1+1/u}{r}\frac{d r}{d u}=-\frac{\ell-2}{2u+\ell},
\end{equation}
and we get
\begin{equation}
g_{tt}=-e^{2\alpha}=-C^2 (2u+\ell)^{-(\ell-2)},
\end{equation}
where $C$ is a constant.
In the coordinate $(t,u,\theta,\phi)$, the $uu-$component of the metric is
\begin{equation}
g_{uu}=e^{2\beta}r'(u)^2=(\ell-2)^2R_0^2\frac{(2u+\ell)^{\ell-2}}{(1+u)^4}.
\end{equation}

Now we try to find a new coordinate system $(t,\rho,\theta,\phi)$, satisfying (1) $g_{tt}\cdot g_{\rho\rho}=-1$, (2) $\rho\rightarrow r$ when $r\rightarrow \infty$.
Condition (1) implies that 
\begin{equation}
\frac{(\ell-2)^2C^2R_0^2}{(1+u)^4}u'(\rho)^2=1,
\end{equation}
which gives
% and combining condition (2) we have
\begin{equation}
u=-1+\frac{(\ell-2) C R_0}{\rho}.
\end{equation}
Combining condition (2) and \eq{eq:solu}, we have
\begin{equation}
C=(\ell-2)^{\ell/2-1}.
\end{equation}
We finally arrive at the result that, in the coordinate system $(t,\rho,\theta,\phi)$, the metric can be expressed as~\footnote{This solution is very similar to the Janis-Newman-Winicour metric~\cite{Janis:1968zz,Virbhadra:1995iy}: \(ds^2=-(1-R_s/r)^\gamma dt^2+(1-R_s/r)^{-\gamma}dr^2+(1-R_s/r)^{1-\gamma}r^2 d\Omega^2\), with \(\gamma=2-\ell\), but a major difference is an extra \(1-R_s/r\) in front of the angular part.}
\begin{equation}
   g_{\mu\nu}=\mathrm{diag}\left(-A(\rho),A(\rho)^{-1},R(\rho)^2,R(\rho)^2\sin^2\theta\right), \label{eq:sol2}
\end{equation}
where 
\begin{equation}
\begin{aligned}
A(\rho)&=\left(1-\frac{R_s}{\rho}\right)^{2-\ell},\\
R(\rho)^2&=\left(1-\frac{R_s}{\rho}\right)^{\ell}\rho^2,
\end{aligned}
\end{equation}
and $R_s$ is a redefinition of a combination of $\ell$ and $R_0$.
For $B_\mu$, the result is
\begin{equation}
    b_t(r)=\sqrt{\frac{2}{\kappa}}\left(1-\frac{R_s}{\rho}\right)^{1-\ell/2}.
\end{equation}
The ADM mass can be read off from the metric as
\begin{equation}
    M_{\rm ADM}=\frac{(2-\ell)R_s}{2 G}.\label{eq:admmass}
\end{equation}
The Kretschmann scalar of this solution is
\begin{align}
K&=R_{\mu\nu\rho\sigma}R^{\mu\nu\rho\sigma}\nonumber\\
&=\frac{(\ell-2)^2R_s^2}{4(\rho-R_s)^{2\ell}\rho^{8-2\ell}}\times\Big(48\rho^2 +32 (\ell-3)R_s\rho\nonumber \\
&\ \ \ \,+(56-36\ell+7\ell^2)R_s^2\Big),   
\end{align}
so when $\ell>0$ and $\ell\neq 2$, $\rho=R_s$ is its singularity, and when $\ell<4$ and $\ell\neq 2$, $\rho=0$ is its singularity.
When $\ell=2$, this solution degenerates into the flat space-time, just the same as the case discussed in Sec.~\ref{sec:Constantb}.
When $\ell\rightarrow 0$, we return to the extremal Reissner–Nordstr{\"o}m black hole solution discussed above.

The Einstein tensor for the solution \eq{eq:sol2} is
\begin{align}
    G^\mu_\nu=\mathrm{diag}\big(&(\ell^2-4)A,-(\ell-2)^2A,\nonumber \\
    &(\ell-2)^2A,(\ell-2)^2A\big),
\end{align}
where 
\begin{equation}
    A=\frac{R_s^2}{4\rho^4}\left(1-\frac{R_s}{\rho}\right)^{-\ell}.
\end{equation}
Still, when $0\leq \ell\leq2$, the metric satisfies the DEC, WEC, NEC, NDEC, and SEC.
It is noteworthy that this is also a sufficient condition for the ADM mass \eq{eq:admmass} to be non-negative.

From the solutions \eq{eq:sol1} and \eq{eq:sol2}, we know that there is no Schwarzschild-like solution when the bumblebee field obtains a time-like VEV as $b=\sqrt{2/\kappa}$ \add{in the Class I cases}.

%notice that $\ell\neq 2$,

\section{Instability, Naked Singularities, and Photon Spheres}\label{sec:5}

\subsection{The stability of the solutions with VEV \texorpdfstring{\(b=\sqrt{2/\kappa}\)}{}\label{sec:5-1}}

Here we argue that the solutions with the VEV \(b=\sqrt{2/\kappa}\) discussed in Sec.~\ref{sec:4} are generally unstable, and thus complete one of our main results: \emph{there is no non-trivial Class I static spherical solution for bumblebee gravity with time-like VEVs}.

First, there is a necessary condition for the stability of the bumblebee model: the Hamiltonian is bounded from below~\cite{Bluhm:2008yt}, and thus the functional form of the potential is restricted.
To our knowledge, stable models are only known for \(V\propto X\equiv B_\mu B^\mu+sb^2\) and \(V\propto M(n,2,X/\mu^2)-1\)~(\(n\ge 3\) for time-like VEVs), where \(\mu^2\) is an energy scale of the theory and \(M(\alpha,\beta,z)\) is the confluent hypergeometric (Kummer's) function~\cite{Altschul:2005mu,abramowitz2006handbook}.
However both types of potentials only satisfy the condition \(V(0)\) while violating \(V'(0)=0\), for which one cannot solve the equations of motion by assuming a fixed vacuum expectation value of \(B_\mu\).
On the other hand, the conditions in \eq{vacuumcondition} can most easily be realized by polynomial functions \(V\propto X^n,\ n>1\), and it is shown that in general there is no lower bound on the Hamiltonian~\cite{Bluhm:2008yt,Carroll:2009em,Bailey:2025oun} for such kinds of potentials, making the models not stable.
\add{
Specifically, we consider a quadratic potential and the flat spacetime, \(V=\frac{\lambda}{2}(X+sb^2)^2\), which is enough for this work and the discussion of stability.
As is shown in Ref.~\cite{Bailey:2025oun}, the Hamiltonian reads
\begin{equation}
    H=\int d^3x\left(\frac{1}{2}\left(\Pi^i\right)^2+\frac{1}{4}B_{ij}B^{ij}+V+2V'B_0^2\right),\label{eq:hamiltonian}
\end{equation}
where \(\Pi^i\) is the conjugate momentum.
Then the Hamiltonian yields an effective potential which can be written as~\cite{Bailey:2025oun}
\begin{equation}
    \frac{1}{\Lambda^4}\mathcal{H}_V=\frac{V}{\Lambda^4}+2\frac{V'}{\Lambda^2}f^2,\label{eq:potentialHV}
\end{equation}
where \(\Lambda\) is some constant of dimensions of mass representing the scale of the theory, and \(f=B_0/\Lambda\).
Explicitly, \eq{eq:potentialHV} reads
\begin{align}
    \frac{1}{\Lambda^4}\mathcal{H}_V=\frac{\lambda}{2}&\left(-f^2+u^2+s\left(\frac{b}{\Lambda}\right)^2\right)\nonumber\\ \times&\left(3f^2+u^2+s \left(\frac{b}{\Lambda}\right)^2\right),\label{eq:potential}
\end{align}
with \(u=||\vec{B}||/\Lambda\).
Since we focus on time-like VEVs, that is, \(f^2>u^2\), it is not hard to see from \eq{eq:potential} that the Hamiltonian is not bounded from below, indicating that all states tend to decay.
The analysis is completely consistent with that of Ref.~\cite{Bailey:2025oun}.
}
This observation then implies the instability of the solutions with \(b=\sqrt{2/\kappa}\).
\add{To be a bit tangential, the boundedness of the Hamiltonian is a rather strict restriction on the forms of the potential, as is partially addressed in Ref.~\cite{Bailey:2025oun}.
At present, only the aforementioned Proca-type potential and the Hypergeometric potentials are proven to be stable, at least under certain conditions.
}

Second, the VEV condition \(b=\sqrt{2/\kappa}\) is such a strict requirement that any deviation, regardless of how small it is, will ultimately cause the non-trivial solutions to degenerate into the Minkowski space-time.
This is evident if we observe that \eq{maineq} forces \(\alpha'(r)\) to be zero once \(\kappa b^2\ne 2\) except at \(r=0\).
Although the seemingly numerical coincidence is maintained once we set \(b=\sqrt{2/\kappa}\) at the classical level, quantum corrections would unavoidably shift the VEV of the bumblebee field \(B_\mu\), and lead the space-time solutions to decay into the flat space-time.
If we focus on the bumblebee field sector, similar to the situation in quantum field theories, loop corrections in general could modify the form of the potential \(V\to V_{eff}\), for which the location of the minimum can be different from \(0\) as in \eq{vacuumcondition}~\cite{Coleman:1973jx}.
An immediate consequence is that the VEV gets modified correspondingly \(b\to b_{eff}\ne\sqrt{2/\kappa}\), and hence no non-trivial solutions are admitted.
Put differently, the non-trivial solutions with \(b=\sqrt{2/\kappa}\) are not stable once quantum fluctuations are taken into account.
Consequently, we conclude again that the solutions with non-zero time-like VEVs in this work are unstable.
Of course, this argument fails if, because of some unknown mechanisms of the theory of QG, \(b\) is \emph{fine-tuned} to be exactly \(\sqrt{2/\kappa}\), which holds to any order of quantum corrections; or, the gravity coupling constant \(\kappa\) would also change by QG corrections such that \(b_{eff}=\sqrt{2/\kappa_{eff}}\) throughout.
However, even without full understanding of the QG theories, we believe that these mechanisms preventing the non-trivial solutions from decay are quite impossible, since such fine-tuning seems to be extremely artificial.

In summary, even each of the two arguments above may be invalid once a complete theory of quantum gravity is taken into account, they, put together, still strongly imply the non-existence of a non-trivial solution with any value of \(b\) in the case of time-like VEVs studied in our work.

\subsection{Violation of the \emph{weak cosmic censorship conjecture}}

The weak cosmic censorship conjecture~(WCCC), proposed by Roger Penrose~\cite{Penrose:1969pc}, serves as a method to save the predictability of general relativity~\cite{Singh:1997wa}.
Obviously, our solutions in this work provide explicit counterexamples to this conjecture, as we will show later.
% In fact, the solution \eq{eq:sol1} and the corresponding Kretschmann scalar \eq{eq:kretschmann1} indicate that there is a singularity located at \(r=0\), but no horizon exists in this case, meaning the existence of a naked singularity, and the violation of the WCCC consequently. 
Nonetheless, we have argued that the solutions in Sec.~\ref{sec:4} are unstable; therefore, the violation of the WCCC appears to be harmless, since the solutions would finally decay into the flat space-time.
And the natural expectation of the value of \(\ell\) also supports this argument.

Regardless of the stability of the solutions, we analyze the properties of the naked singularities in more detail.
The question is whether the naked singularities in our solutions are global, where a singularity is globally naked if there is a future-directed causal curve with one end {\it on the singularity} and the other end on the future null infinity~\cite{Virbhadra:1995iy,Virbhadra:1996cz}.
Although the best way to secure the prediction power is cover the singularity behind an event horizon, a singularity could still be allowed if it is not globally naked or even not locally naked.
Let us first consider the solution \eq{eq:sol1} and only focus on the radial part of this metric. Then, for a outward-directed photon, its geodesic motion is governed by
\begin{equation}
    0=-\left(\frac{r}{R_0}\right)^{\frac{2(2-\ell)}{\ell}}dt^2+\frac{4}{\ell^2}dr^2,
\end{equation}
where the parameter \(\ell\) is restricted by \(0\le\ell\le 2\).
A straightforward calculation gives the propagating time taken by a photon which is emitted at \(r=0\) moving to \(r=R>0\) as
\begin{equation}
    \Delta t=\lim_{\epsilon\to 0^+}\int_\epsilon^R \frac{2}{\ell}\left(\frac{r}{R_0}\right)^\frac{\ell-2}{\ell}=\lim_{\epsilon\to 0^+}\left.\frac{r (r/R_0)^{\frac{l-2}{\ell}}}{\ell-1}\right|^R_\epsilon .
\end{equation}
If \(\ell>1\) then \(\Delta t\) is finite, and the singularity is globally naked;
while for \(\ell <1\) we have \(\Delta t=\infty\), making the singularity even not naked locally.
Next, we consider the radial motion of a photon in the spacetime determined by \eq{eq:sol2}.
Similarly, considering a photon emitted outwards at \(r=R_s\) and reach an observer at \(r=R>R_s\), we can write the propagating time as
\begin{equation}
    \begin{aligned}
        \Delta t=\lim_{\epsilon\to 0^+}\int_{R_s+\epsilon}^R\left(1-\frac{R_s}{\rho}\right)^{\ell-2}d\rho,
    \end{aligned}
\end{equation}
where we again require \(0\le\ell\le 2\) such that energy conditions are satisfied.
By estimating this integral, we come to the same conclusion that for \(\ell>1\) there is a globally naked singularity, while for \(\ell<1\) the singularity is not naked, even locally.

A few comments are in order.
The parameter \(\ell=\xi b^2\) plays a very important role in the analysis of the singularities.
As we can see, regardless of which solutions we are considering, \(\ell>1\) always means globally visible singularities, giving counterexamples of the weak cosmic censorship conjecture.
On the other hand, however, if \(\ell<1\), even though the singularities are not hidden by event horizons, our ability to make predictions is still maintained.
Meanwhile, no energy condition is violated for both cases.
But which case is preferred?
Let us come back to the action \eq{action}, which contains not only the usual gravity and matter contributions, but also a vector field sector {\it non-minimally} coupled to gravity.
The coupling constant of the non-minimally coupled term can be read off as \(\xi/2\kappa=\ell/4\), where we have used \(\ell=\xi b^2=2\xi/\kappa\).
Since \(\ell\) is effectively the degree to which the vector field is coupled to gravity non-minimally, and generally we expect that non-minimal couplings are small, \(0\sim \ell<1\) is naturally favored~\footnote{Besides, in Ref.~\cite{Casana:2017jkc}, the authors showed that this parameter has an upper-bound of \(\ell< 10^{-13}\). Although the solution in Ref.~\cite{Casana:2017jkc} was obtained with a space-like background VEV, the parameter \(\ell\) in both cases should have the same order of magnitude.}.
Consequently, the singularities in our solutions bring no problem to the power of making predictions.

\add{
It is also beneficial to examine the geodesic completeness next. Specifically, let us consider radial time-like geodesics
 \begin{equation}
     -d\tau^2=g_{tt}dt^2+g_{rr}dr^2,\label{eq:geodesic}
 \end{equation}
 where $\tau$ is the proper time along the geodesics. Since the solutions to be examined always have a time-like Killing vector \(K^\mu=(\partial_t)^\mu=\delta^\mu_{\ 0}\), we can construct a constant as
 \begin{equation}
     E=-K_\mu \frac{dx^\mu}{d\tau}=-g_{tt}\frac{dt}{d\tau}.\label{eq:constant}
 \end{equation}
 Combining \eq{eq:geodesic} and \eq{eq:constant}, we obtain
 \begin{equation}
     \frac{d\tau}{dr}=-\frac{\sqrt{-g_{tt}g_{rr}}}{\sqrt{E^2+g_{tt}}}.\label{eq:propertime}
 \end{equation}
 First we check the geodesic completeness for metric~\eqref{eq:sol1}, for which \eq{eq:propertime} leads to
 \begin{equation}
     \tau=-\frac{2}{\ell}\int_{R}^{0}\frac{dr}{\sqrt{E^2(r/R_0)^{2(\ell-2)/\ell}-1}},\label{eq:propertime1}
 \end{equation}
 where the arbitrary and finite radius \(R>0\).
 We do not need to explicitly perform the integration, and a qualitative analysis shows that this solution is geodesic \emph{incomplete}. Actually, there is a critical radius for each choice of \(E\): \(r_c=E^{\ell/(2-\ell)}R_0\). If \(R<r_c\), given \(0<\ell<2\), it is straightforward to show that the integration \eqref{eq:propertime1} is indeed finite, indicating geodesic incompleteness.
 From the analysis of time-like geodesics, a rather weird but intriguing behavior of this spacetime emerges: to go to infinity \(r=+\infty\), a massive test particle should have infinitely large energy, which can be seen from \(r_c=E^{\ell/(2-\ell)}R_0\). In this sense, the spacetime is even \emph{closed} for massive particles, and interestingly, this spacetime appears to exhibit a nature of \emph{confinement}.
 Similarly, for the metric~\eqref{eq:sol2}, we have
 \begin{align}
     \tau&=-\int_R^{R_s}\frac{dr}{\sqrt{E^2-(1-R_s/r)^{2-\ell}}}\nonumber\\
     &=-R_s\int_{R/R_s}^1\frac{dr}{\sqrt{E^2-(1-1/r)^{2-\ell}}},
 \end{align}
 noting that the singularity is located at \(r=R_s\) rather than \(r=0\).
 Since again we require \(0<\ell<2\), the proper time could be finite with suitable choices of the ratio \(R/R_s\) and the ``energy'' \(E\).
 As a result, this solution is also geodesic \emph{incomplete}.
 But in this spacetime, a massive test particle can arrive at infinity as long as \(E>1\), very similar to the case in the Schwarzschild spacetime.
}

\subsection{Photon spheres and observations}

Although the solutions do not have clear physical pictures, {\it i.e.,} whether they are merely mathematically acceptable solutions, or are results of some physical processes such as gravitational collapse, it is helpful to study their aspects of observations.

Here we would like to analyze the photon spheres of solutions \eq{eq:sol1} and \eq{eq:sol2}.
A photon sphere is defined to be a nowhere-spacelike hypersurface, such that any photon whose 4-velocity is initially tangent to the surface will always remain in the surface if not altered~\cite{Claudel:2000yi}.
More intuitively, as a light ray approaches the photon sphere, the Einstein bending angle tends to be infinitely large~\cite{Virbhadra:1999nm,Virbhadra:2002ju}.
For a Schwarzschild black hole, the gravitational lensing would cause {\it relativistic images} due to photon spheres, besides the ordinary primary and secondary images~\cite{Virbhadra:1999nm}.
Furthermore, the existence or non-existence of photon spheres around a singularity~(naked or not) has important implications for gravitational lensing, which can be compared to realistic observations.
In the literature, a singularity with or without photon spheres is termed a  {\it weakly} or {\it strongly naked singularity}~(WNS or SNS) respectively.
To determine the photon spheres, considering a general static spherically symmetric spacetime described by~\cite{Virbhadra:2002ju,Virbhadra:2007kw}
\begin{equation}
    ds^2=-B(r)dt^2+A(r)dr^2+D(r)r^2 d\Omega^2,
\end{equation}
we can obtain the photon sphere by solving
\begin{equation}
    2D(r)B(r)+rD'(r)B(r)-rB'(r)D(r)=0,\label{eq:photonsphere}
\end{equation}
where only constant \(r\) solutions are considered.
As an example, the photon sphere obtained by solving \eq{eq:photonsphere} for a Schwarzschild metric is \(r=3M\), which is exactly the desired result.

First, we examine the spacetime expressed by \eq{eq:sol1}.
Combining \eq{eq:sol1} with \eq{eq:photonsphere}, we obtain
\begin{equation}
    \frac{4(1-\ell)r^{4/\ell-2}}{\ell R_0^{2-4/\ell}}=0,
\end{equation}
which has no reasonable solution since \(0<\ell<2\).
Thus, we conclude that the singularity of \eq{eq:sol1} is {\it strongly naked}.
Then we perform the same analysis for the spacetime \eq{eq:sol2}, and \eq{eq:photonsphere} gives
\begin{equation}
    \frac{2(r-R_s)\left(r-(2-\ell)R_s\right)}{r^2}=0,
\end{equation}
leading to \(r=R_s\) or \(r=(2-\ell)R_s\).
\replace{As a result, there is at least one photon sphere for the singularity, and hence it is {\it weakly naked}.
Furthermore, as we have argued, to make the singularity globally (even locally) {\it invisible}, \(\ell\) should be less than 1, so the outermost photon sphere is located at \(r=(2-\ell)R_s\).}{
But we have already shown that \(R_s\) is actually a singularity, so there is only one photon sphere located at \(r=(2-\ell)R_s\). Furthermore, we also argue that, to make the singularity globally~(even locally) \emph{invisible}, \(\ell\) should be less than \(1\), making \((2-\ell)R_s>R_s\), and rendering the singularity \emph{weakly naked}.
Put differently, requiring the singularity to be weakly naked in this case is equivalent to having a photon sphere surrounding the singularity.
}
According to Refs.~\cite{Virbhadra:2002ju,Virbhadra:2007kw}, a Schwarzschild black hole and a WNS share the \replace{smae}{same} qualitative gravitational lensing features, and the quantitative values are also very close; while for a SNS quite different qualitative features emerge, distinguishable from a Schwarzschild black hole.
As a result, the type of spacetime expressed by \eq{eq:sol2} cannot be distinguished from a Schwarzschild black hole by gravitational lensing, but the gravitational lensing can be used to distinguish \eq{eq:sol1} from a Schwarzschild black hole.

% \subsection{Comparison to Janis-Newman-Winicour and Joshi-Malafarina-Narayan spacetimes}

\section{Summary}%\label{sec:6}

In this work, we study the \add{Class I (\(b_r\equiv0\))} static spherical vacuum solution for bumblebee gravity with time-like VEVs as $b_\mu=(b_t(r),0,0,0)$ such that \(b^\mu b_\mu=-b^2=\text{const}\). 
\add{Different from the Class II (\(R_{rr}\equiv0\)) solutions in Ref.~\cite{Zhu:2025fiy},}
\replace{The}{the} results suggest that for general VEVs, there is no consistent non-trivial solution if $b\neq\sqrt{2/\kappa}$.
With $b=\sqrt{2/\kappa}$, we find two non-trivial solutions as \eq{eq:sol1} and \eq{eq:sol2}, none of which is Schwarzschild-like, but the second one shares some similarity with Janis-Newman-Winicour spacetime~\cite{Janis:1968zz}.
The solutions both exhibit naked singularities, but we argue that these solutions are unstable, which seems to make the singularities harmless.
Nevertheless, we show that in the minimal coupling limit, {\it i.e.,} \(\ell\propto\xi b^2\ll 1\),
the singularities are not globally visible, and they are even locally invisible, thus the weak cosmic censorship conjecture appears to be still valid.
Then we analyze the photon spheres and find that \eq{eq:sol1} has no photon spheres, while \eq{eq:sol2} has a photon sphere outside the singularities.
As a result, the gravitational lensing features of \eq{eq:sol1} could be very different from those of a Schwarzschild black hole; however, by only utilizing gravitational lensing features, \eq{eq:sol2} cannot be distinguished from a Schwarzschild black hole.
% Since there are no Schwarzschild-like solutions for bumblebee gravity with time-like VEVs, and only the flat space-time solution is stable, we probably do not need to consider the case of time-like VEVs when related to real physics in future researches.
Finally, the newly obtained solutions give rise to further questions, such as what is the reason for the existence of a solution when $b=\sqrt{2/\kappa}\simeq E_{Pl}$, and we hope future studies could shed light on the questions.

% \add{
% As a final remark, this work focuses mainly on the theoretical aspects of the solutions of the bumblebee model with time-like VEVs, however it would be beneficial to calculate the quasinormal mode and greybody factor~\cite{Kanzi:2021cbg}, which would strengthen the observational relevance and allow direct comparison with real observations, {\it e.g.,} EHT data from M87* and Sgr~A*.
% Hence we would like to perform the analysis in the future, since it is beyond the scope of this work.
% }
\add{
As a final remark, this study has primarily examined the theoretical properties of the solutions arising in the bumblebee model with time-like VEVs. A distinctive feature of our results is that the geometries obtained correspond to naked singularity solutions rather than conventional black holes. This distinction suggests that their dynamical and observational signatures—such as quasinormal modes and greybody factors—may differ substantially from those typically associated with black-hole spacetimes~\cite{Chowdhury:2020rfj}. A systematic computation of these quantities~\cite{Kanzi:2021cbg, Chowdhury:2020rfj} would therefore deepen our understanding of the physical behavior of these solutions and, more broadly, enrich the study of naked singularities by providing a new and explicitly solvable theoretical example.
In addition to dynamical probes, it would also be worthwhile to explore the implications of these solutions for gravitational lensing and the formation of shadow-like structures. Since the causal and geometric properties of naked singularity spacetimes differ markedly from those of black holes, their lensing patterns and shadow profiles may exhibit qualitatively new features, opening up novel possibilities for observational discrimination~\cite{Virbhadra:2002ju,Virbhadra:2007kw}. 
Such analyses could facilitate more direct comparison with astrophysical data, including the Event Horizon Telescope observations of M87* and Sgr~A*. While these extensions would significantly strengthen the observational relevance of our results, they lie beyond the scope of the present work and are left for future investigation.
}

\section*{Declaration of competing interest}

The authors declare that they have no known competing financial interests or personal relationships that could have appeared to influence the work reported in this paper.

\section*{Acknowledgements}
The authors are grateful for the valuable discussions with Hanlin Song and Peixiang Ji.
We also thank Kumar Shwetketu Virbhadra for his valuable correspondence.
This work was supported in part by the National Natural Science Foundation of China~(NSFC) under Grant No.~12547101. HL was also supported by the start-up fund of Chongqing University under No.~0233005203009, and JZ was supported by the start-up fund of Chongqing University under No.~0233005203006.

\bibliographystyle{elsarticle-num}
\bibliography{refs}
% \bibliography{refs,refs2}
%\bibliography{refs2}

\end{document}